\pdfoutput=1
\documentclass[preprint,5p,times]{elsarticle}
\usepackage[T1]{fontenc}
\usepackage[utf8]{inputenc}
\usepackage{graphicx}
\usepackage{textcomp}
\usepackage{natbib}
\usepackage{units}

\journal{NIM A}

\begin{document}
\begin{frontmatter}
\title{Monte-Carlo Simulations for the optimisation of a TOF-MIEZE Instrument}
\author[frm,e21]{T. Weber}
\author[frm,e21]{G. Brandl}
\author[frm,e21]{R. Georgii}
\ead{Robert.Georgii@frm2.tum.de}
\author[frm,e21]{W. Häußler}
\author[frm]{S. Weichselbaumer}
\author[e21]{P. Böni}
\address[frm]{Forschungsneutronenquelle Heinz Maier-Leibnitz,
  Technische Universität München, Lichtenbergstr.\ 1, 85748 Garching, Germany}
\address[e21]{Physik Department E21,  Technische Universität München,
  James-Franck-Str., 85748 Garching, Germany}

\date{\today}

\begin{abstract}
  The MIEZE (Modulation of Intensity with Zero Effort) technique
  is a variant of neutron resonance spin echo (NRSE),
  which has proven to be a unique neutron scattering technique for
  measuring with high energy resolution in magnetic fields. Its limitations in
  terms of flight path differences have already been investigated analytically
  for neutron beams with vanishing divergence. In the present work Monte-Carlo simulations
  for quasi-elastic MIEZE experiments taking into account beam divergence
  as well as the sample dimensions are presented.
  One application of the MIEZE technique could be a dedicated
  NRSE-MIEZE instrument at the European Spallation Source (ESS) in Sweden. The optimisation of
   a particular design based on Montel mirror optics with the help of Monte Carlo simulations will be discussed here in detail.
\end{abstract}
\begin{keyword}
  MIEZE \sep spin echo \sep Monte Carlo Simulation \sep MISANS \sep ESS
\end{keyword}
\end{frontmatter}

\section{Introduction}
One major field of current solid state physics is complex magnetic systems.
Examples are the structure and dynamics of topological spin structures like
skyrmions in metals, semiconductors and ferroelectric insulators
\cite{Muehlbauer2009,Muenzer:09,Adams2012,Seki2012}.  Others are the dynamics of
spin waves in ferromagnets like EuO and Fe \cite{Farago1986100}. Further interest focuses on the slow magnetic dynamics of
magnetic monopoles and Dirac strings in spin ice \cite{Jaubert:2009}.  Generally,
excitations with lifetimes from the picosecond to the microsecond range near quantum
phase transitions are of utmost interest for the understanding of the magnetic
dynamics in complex materials, especially with regard to possible
applications in spintronics.

The standard technique to measure quasi-elastic dynamics in the higher ps and ns time range is Neutron Spin
Echo (NSE) \cite{Mezei:72}.  As a Larmor precession technique
it is sensitive to magnetic fields and can only be used with magnetic
fields at the sample position by applying rather complicated spin manipulation resulting in a loss of
intensity \cite{Farago1986100}.  The complementary Neutron Resonance Spin Echo
(NRSE) method invented by Golub and Gähler \cite{Gaehler:88, Gaehler:92} shares the
sensitivity to magnetic fields or depolarising samples, but can be evolved
straightforwardly into the MIEZE technique \cite{Gaehler:92, Besenboeck:98, Georgii:11},
which is then independent of sample depolarisation effects, as all spin
manipulations are performed before the sample.

As MIEZE is effectively a time-of-flight method it is
sensitive to differences in the length of the neutron flight path.  This drawback has been studied
analytically in reference \cite{Brandl:11} for beams with zero divergence.  But for
analysing MIEZE data and designing MIEZE instruments the effect of
divergence and the influence of the neutron path on the resolution is required.  In this study
we will show that using the Monte Carlo simulation package  McStas 
\cite{Willendrup:04, Lefmann:99} together with specifically written components
for the MIEZE setup, both earlier analytical and experimental results can be reproduced and
furthermore be used to assess the suitability of more elaborate instrument designs.
MIEZE is an excellent technique for high energy resolution measurements:
For small samples spin-echo times of the order of a microsecond are possible with currently available
 components such as RF spin flipper coils and TOF detectors.

\section{MIEZE technique}
As a variant of NRSE, the MIEZE technique omits both $\pi$-flipper coils
after the sample \cite{Gaehler:92}.  The coils before the sample are separated by a distance $L_1$
and are driven at two different frequencies, $\omega_1 < \omega_2$.  After
passing the RF-flippers, which are working in resonant $\pi$-flip mode without bootstrap, the neutron spin phase depends on the flight time and the neutron
velocity $v$. At a distance $L_2$ given by 
\begin{equation}
 \omega_1 \cdot L_1   = \left( \omega_2 - \omega_1 \right) \cdot L_2
  \label{eq:MIEZE_cond}
\end{equation}
the spin modulation of all velocity groups adds up in phase.  Eq.~(\ref{eq:MIEZE_cond}) is called
the MIEZE condition.  The key point of MIEZE is now that the spin modulation can
be converted to an intensity modulation by placing a polariser anywhere after
the second coil, in particular before the sample.  Then a time-resolved detector
is required, which registers a sinusoidally modulated signal of the frequency $\omega_{M}$ 
$I(t) \propto \cos(\omega_M t)$, where  $\omega_M = 2 (\omega_2 - \omega_1)$ is the
MIEZE frequency.

If the phase at the detector is not uniform, which can be caused by inelastic
scattering at the sample or by path length differences after the polariser, the
modulation is damped and the signal can be described by
$I(t) = B + A\cos(\omega_Mt)$.  For negligible path length differences the MIEZE contrast $C = A/B$ is a measure
of the inelasticity of the scattering, which is the MIEZE
equivalent of the measured polarisation in NSE/NRSE.

It can be shown \cite{Keller:02} that with an ideal setup, the contrast is the
cosine Fourier transform of the scattering law $S(q,\omega)$ of the sample
\begin{equation}
  C \left(q, \tau_M \right) = \int\!d\omega\, S(q, \omega) \, \cos(\omega \tau_M),
  \label{eq:MIEZE_cos_ft}
\end{equation}
with the sample placed at the position $L_S$ before the detector.  With $m$ being the mass of the neutron, the MIEZE time $\tau_M$
is given by
\begin{equation}
  \tau_M = \frac{\hbar \omega_M L_S}{m v^3},
  \label{eq:MIEZE_time}
\end{equation}
This is the Fourier time of the instrument, equivalent to the spin-echo time
of NSE/NRSE \cite{Keller:02}.

For a real MIEZE instrument, the contrast of the modulation is reduced by imperfections of
polarisers, coils, detector and by path length differences.  The reduction in contrast $C$ due to different sample
shapes and spin-echo times was studied theoretically in reference \cite{Brandl:11}.
To benchmark the quality of the Monte-Carlo simulation of MIEZE, we reproduce
and verify the previous work in the next section.

\section{MIEZE and Monte-Carlo simulations}

\subsection {\label{sim}Reproduction of previous results}
For the reproduction of the previous results from the reference \cite{Brandl:11}  we used the Monte-Carlo simulation package McStas \cite{Willendrup:04, Lefmann:99}  simulating a standard MIEZE setup without any  additional neutron optics and the sample only. 

The contrast of the MIEZE signal is reduced by several effects of which we discuss the most important ones, namely  the thickness of the absorber in the detectors, the resolution of the detector and the geometry of the sample in more detail. The effect of imperfections  in the NRSE coils on the spin and the dependence on the MIEZE frequency $\omega_{M}$ is not taken into  account here as the coils are simulated as ideal RF spin flippers.  

The thickness $\Delta d$ of the detector modifies the registered contrast  by introducing a phase shift in equation
(\ref{eq:MIEZE_cos_ft})
\begin{equation}
C = \frac{1}{\Delta d} \int d\omega   \int_0^{\Delta d} dx \,
S\left(\omega\right)\cos\left(\omega \tau_{M}+ \omega_{M} x /v \right)
\mbox{.}
\end{equation}
A phase shift of $2\pi$ over the whole thickness of the
detector would
completely destroy the contrast,
whereas a phase shift of $\pi/2$ would not reduce it by more than 10 \%.
For example in the CASCADE detector \cite{Schmidt:2010, Klein:2011jj, Haeussler:2011}  we used in our MIEZE experiments,
where the thickness of the detection planes is
$\Delta d = 2 \,\mu$m, a 10 \% loss of contrast happens at MIEZE
frequencies of
$\omega_{M} = 621$\,MHz, $\omega_{M} = 311$\,MHz and $\omega_{M} =
207$\,MHz for
$\lambda=5$\,\AA, $\lambda=10$\,\AA\, and $\lambda=15$\,\AA, respectively.
For the respective wavelengths this corresponds to a maximum possible
MIEZE time  of
$\tau_{M} = 1 \,\mu$s, $\tau_{M} = 4\,\mu$s and $\tau_{M} = 9 \,\mu$s, 
which is far above the values considered here (see Figures \ref{fig:maxtau} and \ref{fig:qtau}) thus justifying the assumption of a thin detector  in the simulations.

To counter the effect of contrast reduction due to path length differences within the plane of the detector especially at higher MIEZE times $\tau_M$  an area of 1 mm$^2$ is used in the simulation. This is in the order of the spatial resolution of a CASCADE detector.  

As the thickness and spatial resolution of the detector therefore do not contribute significantly  to the contrast reduction, we only have  to consider the reduction factor due to the sample in the following. 

\begin{figure}[t]
  \begin{center}
    \includegraphics[width=0.95\linewidth]{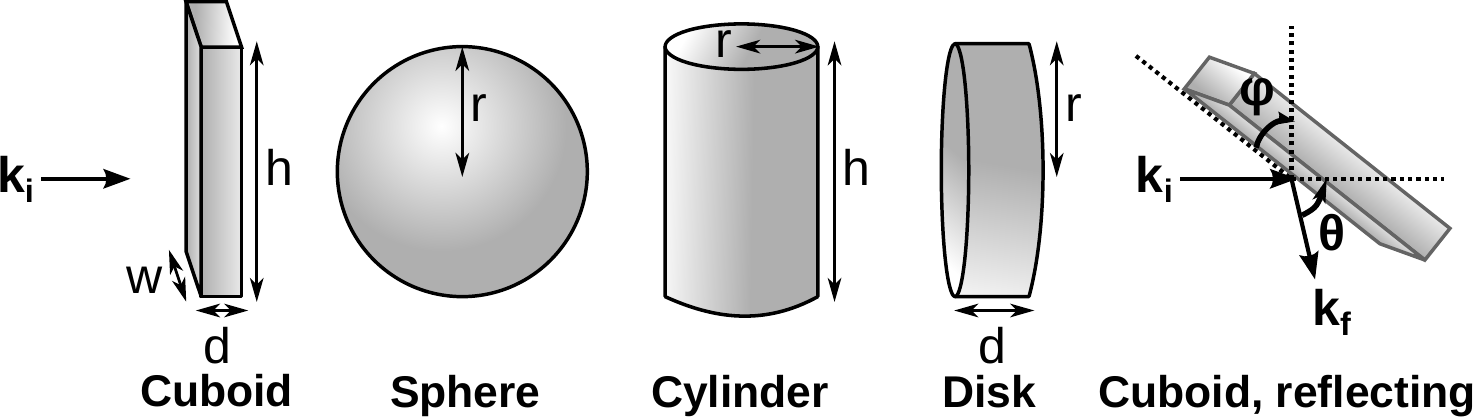} \\[1ex]
    \includegraphics[width=\linewidth]{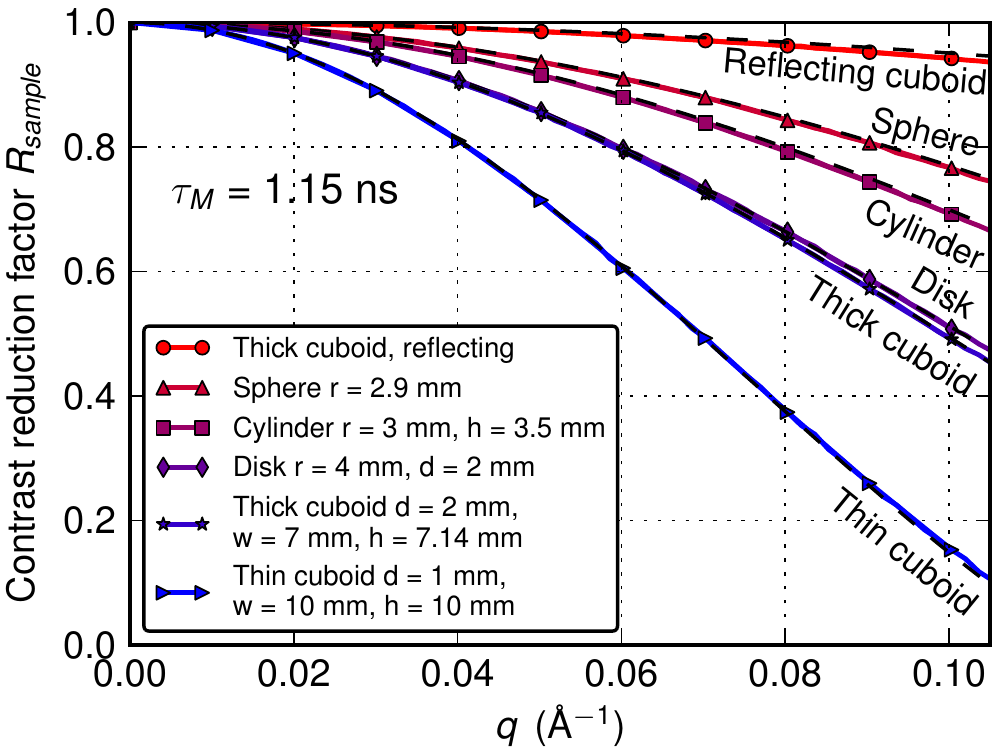}
    \caption{\label{fig:geos}Top: The different sample shapes are shown. In the
    reflecting geometry the sample is rotated at an angle $\phi = \theta/2 - 90^\circ$ with respect to the
scattering angle as indicated in the figure.
    Bottom: Comparison of the sample reduction factor for different
    geometries with constant sample volume $V = 100$ mm$^3$ using
    the analytical results from \cite{Brandl:11} (dashed lines) and
    simulated results (solid lines).  Parameters are as given in
    \cite{Brandl:11}: $\omega_M$ = $2 \pi \cdot 200$\,kHz, corresponding to
    $\tau_M=1.15$\,ns, $L_1=1$\,m, $L_2=2$\,m, $L_s=0.8$\,m, $\lambda=10.4$\,\AA.}
\end{center}
\end{figure}

In the simulations the contrast is normalised to its value at $q = 0$. Vanadium
is used as a purely elastic scatterer. Fig.~\ref{fig:geos} shows a comparison of the simulated and calculated \cite{Brandl:11}
reduction factors due to different sample shapes in fixed and reflecting geometry
(where reflecting is defined as a rotation $\phi = \theta/2 - 90^\circ$ in
the direction of $\theta$, see Fig.~\ref{fig:geos})
For the spherical and cylindrical samples absorption is neglected.
This simplification is necessary because the analytical formulae in
\cite{Brandl:11} do not take into account the additional contrast reduction
caused by inhomogeneous absorption due to sample shape.
The results demonstrate an almost perfect agreement between the analytical and the simulated results.

\begin{figure}[h]
  \begin{center}
    \includegraphics[width=\linewidth]{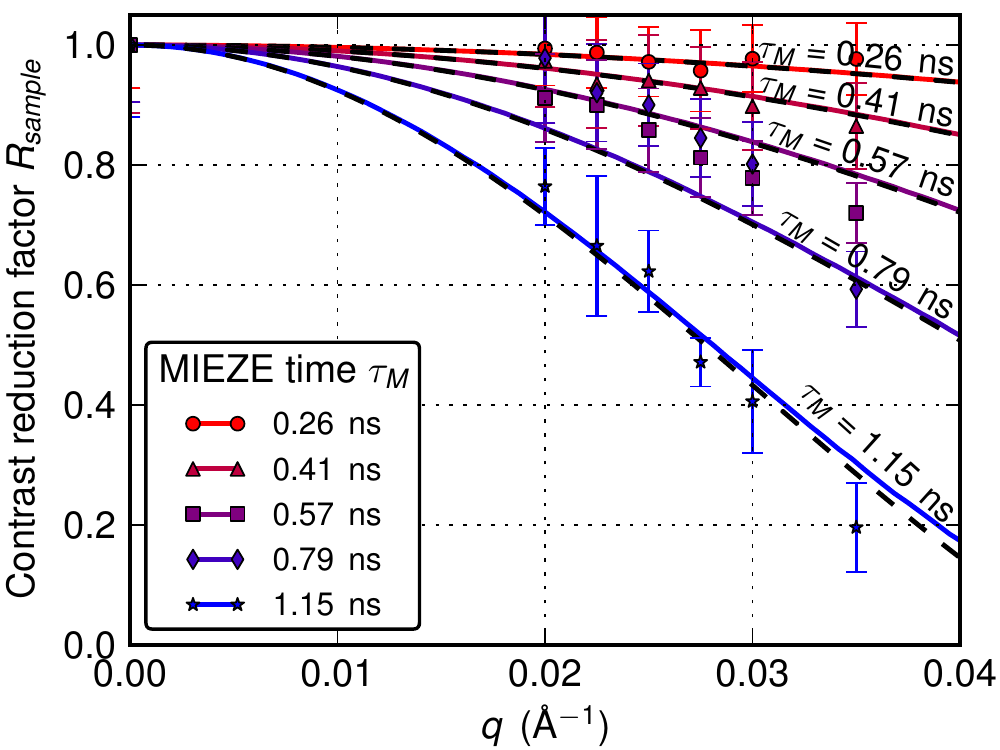}
    \caption{\label{fig:taus} Comparison of the sample reduction factor for a
      cuboid in fixed geometry showing the reduction factor for the simulation
      results (solid lines), the analytical results from \cite{Brandl:11}
      (dashed lines) and experimental (symbols with error bars) results from the instrument MIRA also
      performed by \cite{Brandl:11}.
      Parameters are as given in \cite{Brandl:11}: sample width 25\,mm,
      thickness 5\,mm, $\lambda=10.4$\,\AA, $\omega_M$ ranges from
      $2 \pi \cdot 46$\,kHz to $2 \pi \cdot 200$\,kHz, yielding a MIEZE time
      $\tau_M$ from 0.26\,ns to 1.15\,ns, respectively. Instrument lengths
      are: $L_1=1$\,m, $L_2=2$\,m, $L_s=0.8$\,m.}
  \end{center}
\end{figure}

Fig.~\ref{fig:taus} shows the simulated (solid lines), calculated (dashed lines) and measured (symbols with error bar) reduction factors for a cuboid sample
in fixed geometry for different spin-echo times $\tau_M$. There is an excellent
agreement between simulation (this work), calculation and measurement (from \cite{Brandl:11} ).

Summarising, the effects of contrast reduction due to the geometry of the sample calculated using an analytical method are very well reproduced by the 
McStas simulations using a standard MIEZE setup with no additional neutron optics (compare  Figs. \ref{fig:geos} and \ref{fig:taus} in this paper to  Figs. \ref{fig:maxtau} and \ref{fig:qtau} in \cite{Brandl:11}).

\subsection {Compensation for detector misalignment}
In a real experiment with a total length in the order of several tens of metres  a detector misalignment relative to the coils violating the 
MIEZE condition, will lead to a reduction of MIEZE contrast. We briefly show that this effect can easily
be compensated by adjusting one of the RF frequencies.

A detector that is misaligned by an offset of $\Delta l$ away from $L_2$ does
not fulfill the MIEZE condition (Eq.~\ref{eq:MIEZE_cond}) anymore.  However, the
signal can be recovered by adjusting the frequency of one of the $\pi$-flipper
coils.  This can immediately be seen if we substitute $L_2$ by 
$L_2 + \Delta l$ and $\omega_1$ by $\omega_1 + \Delta \omega_1$ in
eq.~(\ref{eq:MIEZE_cond}) and solve for $\Delta \omega_1$:
\begin{equation}
\Delta\omega_{1}=\left(\omega_{2}-\omega_{1}\right)-\frac{L_{1}\cdot\omega_{2}}{\Delta
l+L_{1}+L_{2}}  \label{eq:freq_scan}
\end{equation}

\begin{figure}[h]
  \begin{center}
    \includegraphics[width=\linewidth]{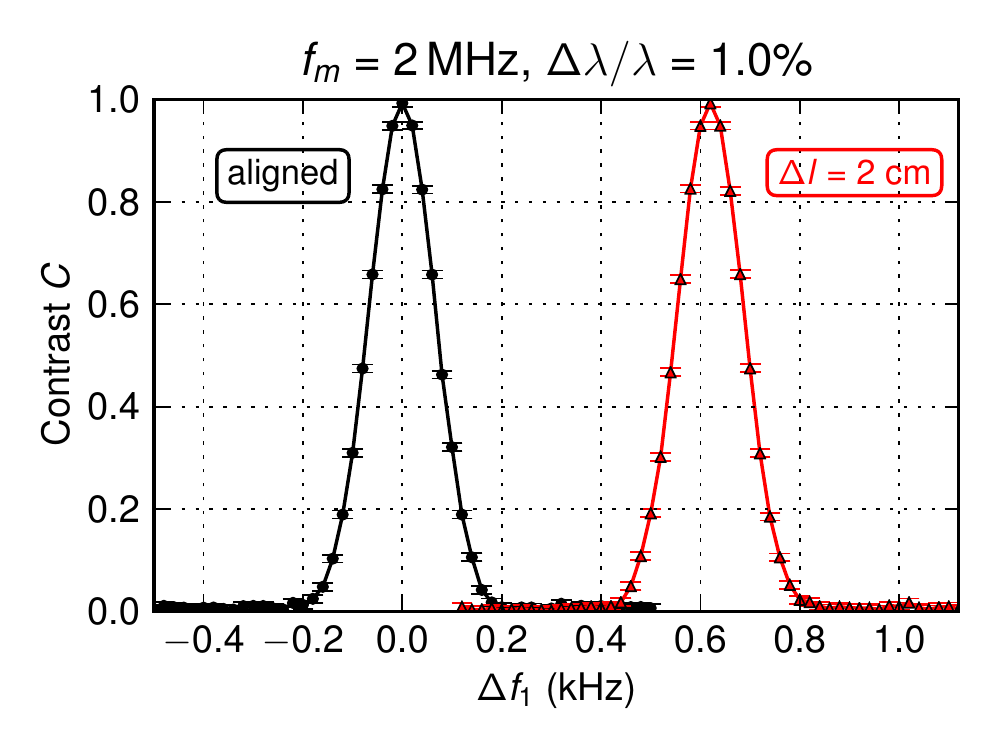}
    \caption{Compensation of a detector offset of 2 cm by adjustment of the
      first spin flipper frequency.  The left curve  shows the signal
      for an aligned detector, the right curve  demonstrates that the
      signal can be recovered for a misaligned detector by changing the
      frequency $f_M= \omega_M/2\pi$ of the second coil according to Eq.~(\ref{eq:freq_scan}). 
      Instrument parameters are: $L_1=16$\,m, $L_2=16$\,m, $L_s=13$\,m, $\lambda=15$\,\AA.
      \label{fig:freq_scan}}
  \end{center}
\end{figure}

According to Eq.~(\ref{eq:freq_scan}), the first spin flipper frequency $\omega_1$  has to
be adjusted by $\Delta \omega_1$ to compensate for a detector
misalignment of $\Delta l$.
The simulations (Fig.~\ref{fig:freq_scan}) show that the signal can be perfectly recovered.

\section{A dedicated NRSE-MIEZE instrument at the ESS}

\begin{figure}[h]
  \begin{center}
    \includegraphics[width=\linewidth]{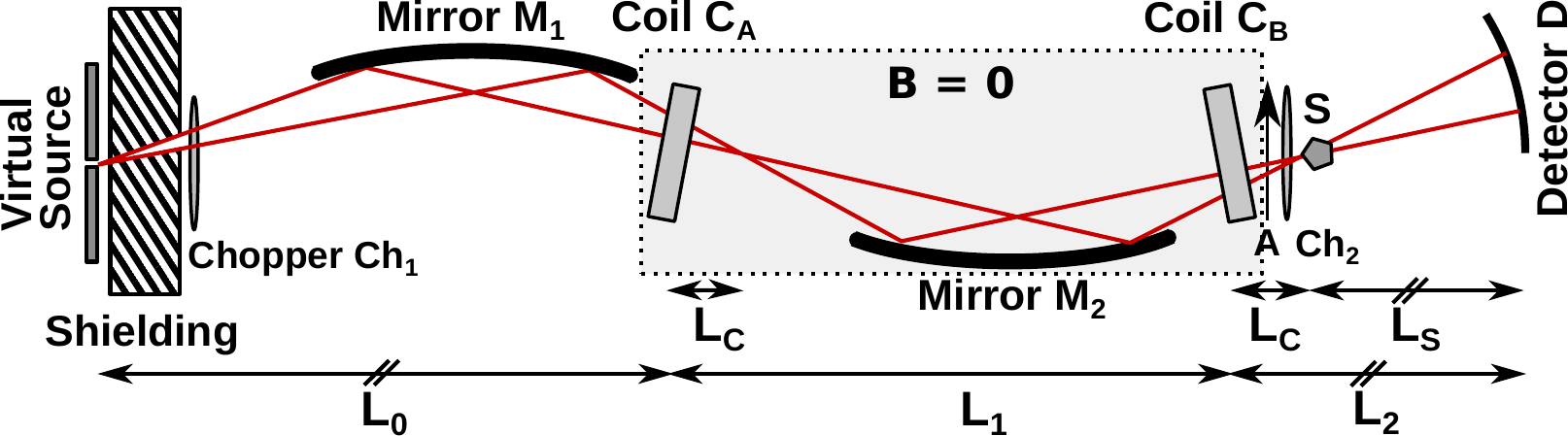}
    \caption{Schematic view of the MIEZE instrument using Montel optics.
      $M_1$ and $M_2$ are reflecting mirrors with a length of 9.6\,m each,
      which is 60\% of the distance $L_1$ between the $\pi$-flipper coils
     $C_A$ and $C_B$, $A$ and $S$ indicate the analyser and the sample, respectively.
      The instrument parameters used in the simulations are:
      Distance from the virtual source with an area of 2 x 2 mm$^2$ to the coil $C_A$: $L_0 = 13$\,m,
      distance $L_1 = 16$\,m,
      distance from coil $C_B$ to the detector: $L_2 = 16$\,m,
      distance from coil $C_B$ to the sample: $L_C = 3$\,m,
      distance from the sample to the detector: $L_s = 13$\,m.
      \label{fig:setup}}
  \end{center}
\end{figure}
\begin{figure}[h]
  \begin{center}
    \includegraphics[width=\linewidth]{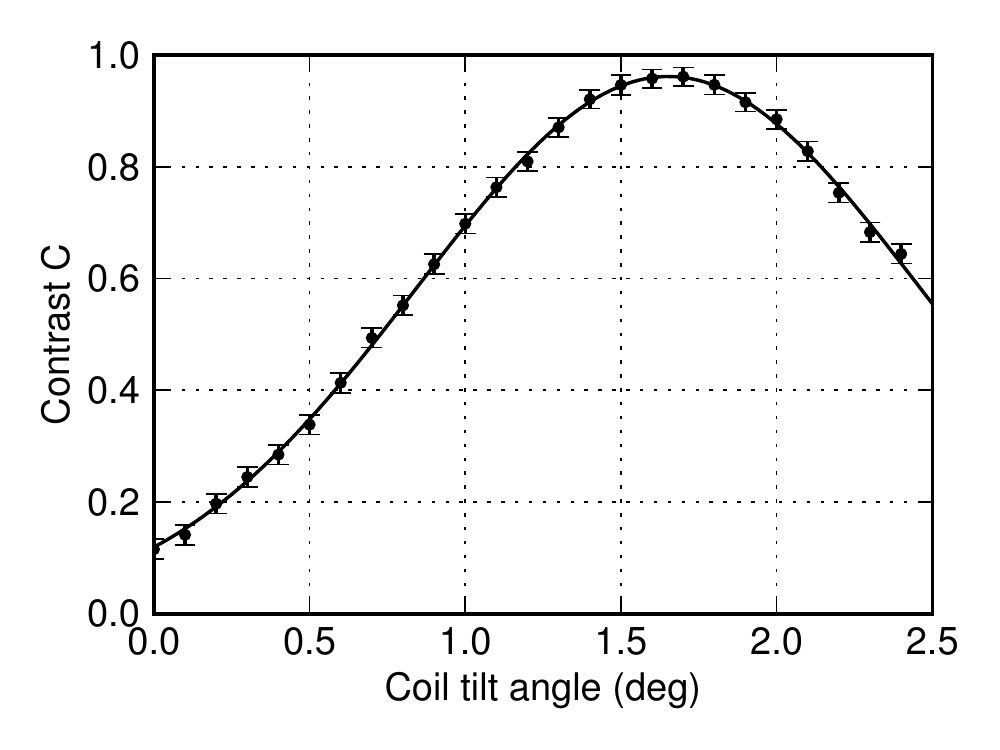}
    \caption{Dependence of the contrast on the tilt of the $\pi$-flipper coils. The optimal coil rotation of 1.65$^0$ is independent
    of $\lambda$.
    \label{fig:rot}}
  \end{center}
\end{figure}

In the following we show how the newly developed simulation tools can be used to demonstrate the feasibility of a focusing MIEZE beam line
for various $q$ and $\tau_M$ parameters.

We will propose a combined MIEZE/NRSE instrument for measuring slow dynamics, in particular
from depolarising samples and samples in extreme environments (magnetic fields, pressure)  at the planned European Spallation Source (ESS) in Lund, Sweden. 
This source will provide neutron pulses with a repetition rate of 14\,Hz, a pulse length of 2.8\,ms and a total power of 5\,MW. The peak 
intensity will reach 30 times the average intensity of the ILL. A sketch of the planned instrument configuration  is shown in
Fig.~\ref{fig:setup}.  It consists of a beam extraction and guide
system based on Montel optics \cite{Montel:57, Stahn:2011,Stahn:2012, Ice:2009}, which allows a
decoupling of the selection of the size and the divergence of the incident beam.  The distance between the focal points of each Montel mirror is projected to be $L_1 = 16$ m. 
The reflecting mirrors are assumed to be 9.6 m long. 
The goal is to transport only those neutrons to the sample that will be useful for scattering from the sample,
i.e. the beam size is equivalent to the sample size and the divergence is such
that it is compatible with the requirements of the experiment. All other neutrons are absorbed before, therefore the background
will be minimised at the sample position and the low radiation load on the transport system for the neutrons will yield a longer life time of the components.

The chopper system will be designed such that it provides neutrons for
wavelength bands $3 \le \lambda \le 9.5$ \AA\ and $9 \le \lambda \le 15.5$ \AA\ by
changing the phase relation between the two choppers.  They are located right
after the biological shielding of the ESS (6\,m downstream the moderator) and in
front of the sample. Therefore, neutrons spanning the whole wavelength band without frame
overlap and using all pulses from the moderator are used.
The lower wavelength band around $6.25$ \AA\ defines the total length of the instrument ($L_{tot} = 45$ m)
from the moderator to the sample position,
which is located 26\,m after the biological shielding (Fig.~\ref{fig:setup}).

\begin{figure}[t]
  \begin{center}
    \includegraphics[width=\linewidth]{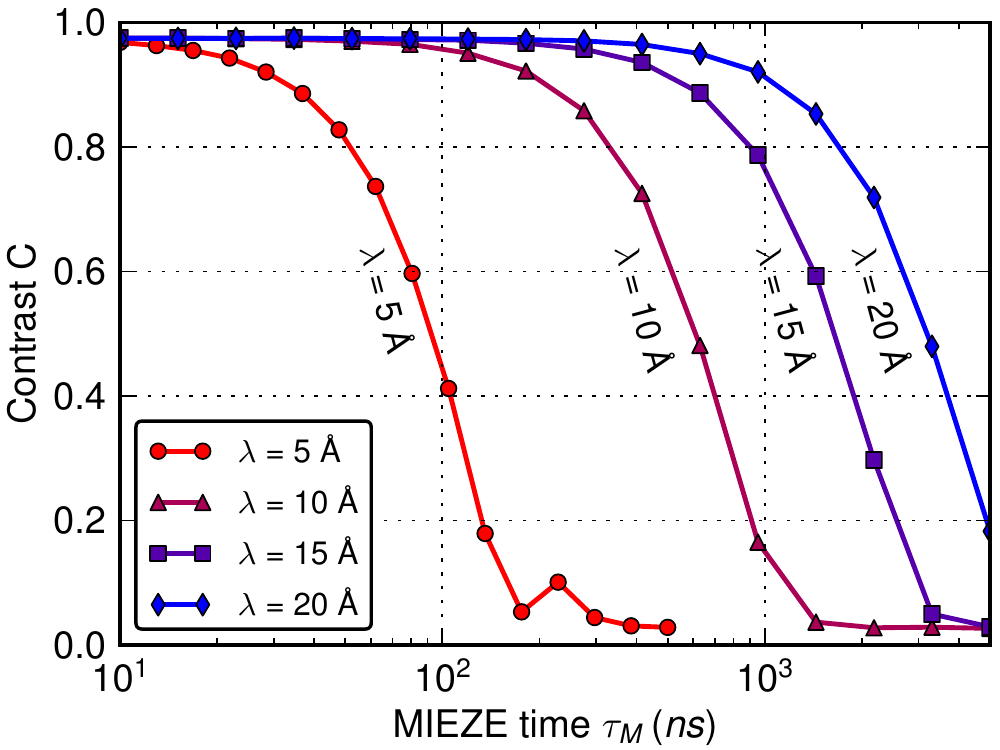}
    \caption{Contrast vs. MIEZE time $\tau_M$ for various wavelengths for the small angle regime without inclusion of the effects of sample shape and gravitation.
      The instrument parameters are: $L_1 = 16$\,m, $L_2 = 16$\,m, $L_s = 13$\,m. 
	  The coating of the Montel optics is $m = 5$ times the critical angle of natural nickel.
      \label{fig:maxtau}}
  \end{center}
\end{figure}

\begin{figure}[h]
  \begin{center}
    \includegraphics[width=\linewidth]{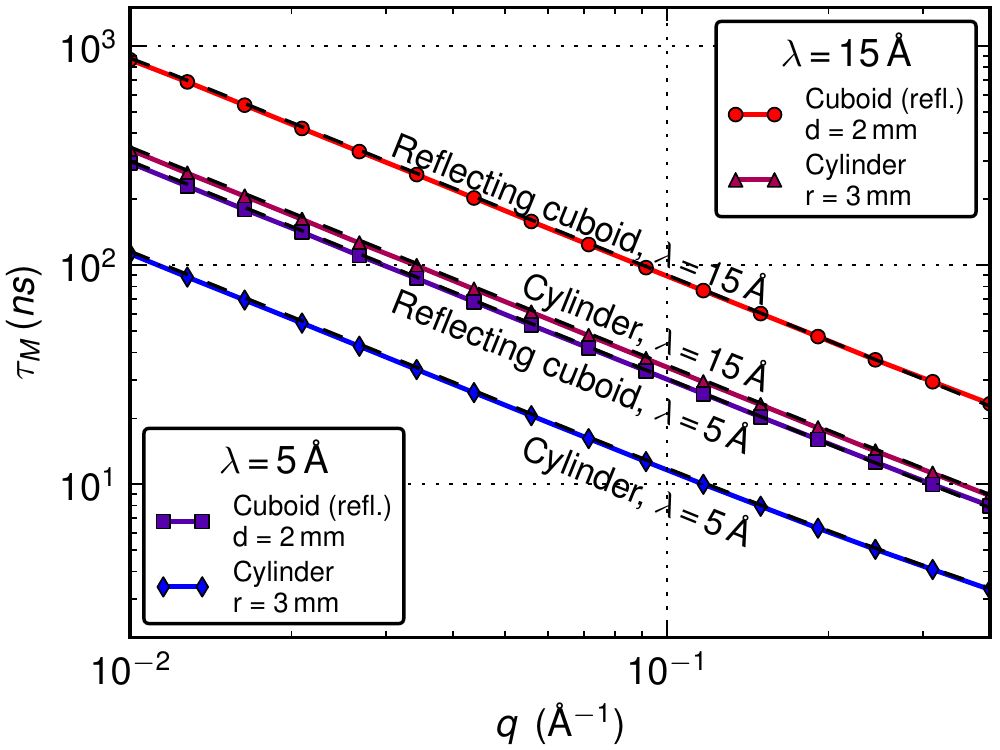}
    \caption{\label{fig:qtau} Time resolution limitations from the sample shape, not including effects of the focussing mirrors. The $R=50\%$ contour
      for a cuboid sample in reflecting geometry with a thickness of 2\,mm (typically used for soft mater) and a cylindrical sample with a radius of 3\, mm (typically used for hard condensed matter and for experiments under extreme conditions) are shown for different wavelengths. Instrument
    parameters are: $L_1=16$\,m, $L_2=16$\,m, $L_s=13$\,m.}
  \end{center}
\end{figure}

The coil system consists of two NRSE coils: one in front of the
sample and one located between the two Montel mirrors of the guide system.
The 16\,m long precession region is shielded by $\mu$-metal for providing a zero field region.  The first Montel
mirror polarises the beam. In front of the sample will be a removable spin analyser.
Running the two coils at different RF-frequencies one obtains a time modulated
intensity at the detector. Two options are possible:

\begin{enumerate}[i.]
\item For the small angle scattering regime, a detector of 1\,m$^2$ with a spatial
resolution of 2\,mm will be placed 10\,m behind the sample allowing a time resolution of the order
of $\mu$s  \, (see Figs.~\ref{fig:maxtau} and ~\ref{fig:qtau}). Such a detector
can be realised for example by combining position-sensitive detectors such as
the CASCADE system \cite{Schmidt:2010, Klein:2011jj, Haeussler:2011}. It also conserves the MIEZE contrast as described above.
\item For larger values of $q$, a set of
NRSE coils behind the sample will correct for the path length differences in the beam.
Using an analyser bank over an angular range of $90^\circ$ and a detector distance
of 2\,m (with a relaxed resolution of 1\,cm), this setup will allow for a $q$-range up
to 3\,\AA$^{-1}$ with a time resolution exceeding 100\,ns. Note that this configuration has not  yet  been simulated as it needs a different simulation method.
\end{enumerate}

The Monte-Carlo simulations as described in section \ref{sim} were now used to simulate the
signal  for the small angle regime of the proposed instrument i.e. option (i) above. The simulation consists of the ESS source, the instrument including the Montel optics, the MIEZE coils and the detector. Perfect MIEZE coils, Montel mirrors  and no gravity were assumed. Note, that the total contrast reduction is obtained by multiplying the individual contrast reductions due to the focusing optics (Fig.~\ref{fig:maxtau})) and the sample shape (Fig. ~\ref{fig:qtau}).

An optimisation of the inclination of the NRSE coils due to the chosen
mirror configuration (see Fig.~\ref{fig:rot}) was performed. The influence of the flight path
differences on the resolution was significantly reduced for an optimal inclination of the coils by 1.65
degrees.  A misplacement of the detectors would degenerate the signal, but can
be compensated by the adjustment of the MIEZE frequency as demonstrated in
Fig.~\ref{fig:freq_scan}.  Using a wavelength of 15 \AA~and the optimal MIEZE frequency, which allow for maximal MIEZE times of up to a microsecond, the
 simulation results show no relevant decrease of the MIEZE contrast at the optimal inclination of the coils  (see Fig.~\ref{fig:maxtau}).
Taking into account the reduction factor due to the sample shape the accessible $q - \tau_M$ space for two typical sample geometries used in the field of magnetism  is plotted in Fig.~\ref{fig:qtau}.  The results demonstrate that the large time resolution
up to  one  microsecond of the instrument can be preserved for selected sample geometries.

In conclusion, we succeeded to prove that Monte-Carlo simulations are
a powerful means for the optimisation of spectrometers using the MIEZE
technique. Choosing a favourable sample geometry can preserve the excellent time resolution of the
instrument. Furthermore, the use of  focusing techniques and Montel optics will lead to
a low background and a tremendously  reduced radiation load on the beam components. Moreover, no mirror optics and no choppers
are required inside the biological shielding thus facilitating the  maintenance 
of a MIEZE beam line. Adjusting the MIEZE frequency will compensate for possible detector misplacements. 
We have shown that path length differences can be effectively compensated by properly
tilting the $\pi$-flippers in a focussing NRSE/MIEZE instrument.   

We have demonstrated that a powerful, low background and low cost MIEZE/NRSE
beam line at the ESS can be realised using beam components that are already available now and are in regular use
at the beam line RESEDA at FRM II \cite{RESEDA} .
The time resolution of a MIEZE/NRSE at large angles will be of the
same order as for NSE SPAN-like instruments \cite{SPAN}. However, going to small q will add the benefit of performing
experiments from depolarising samples and in magnetic fields with a significantly improved time resolution.

\section{Acknowledgements}
This work was funded by the German BMBF under ``Mit\-wirk\-ung der Zentren der
Helmholtz Gemeinschaft und der Technischen Uni\-ver\-sit\"at M\"un\-chen an der
Design-Update Phase der ESS, F\"or\-der\-kenn\-zeichen 05E10WO1.''

\bibliography{mieze}
\end{document}